\documentclass[conference]{IEEEtran}

\usepackage{dependencies}
 \pgfplotsset{compat=1.18}

\begin{document}

\title{\textit{The Danger Within}: Insider Threat Modeling\\ Using Business Process Models}

\author {
  \IEEEauthorblockN {
    Jan von der Assen\IEEEauthorrefmark{1},
    Jasmin Hochuli\IEEEauthorrefmark{1},
    Thomas Grübl\IEEEauthorrefmark{1},
    Burkhard Stiller\IEEEauthorrefmark{1}
  }
  \IEEEauthorblockA {
    \IEEEauthorrefmark{1}Communication Systems Group, Department of Informatics, University of Zurich UZH, CH--8050 Zürich, Switzerland \\{[vonderassen, gruebl, stiller]}@ifi.uzh.ch, jasmin.hochuli@uzh.ch
  }
}
\DeclareRobustCommand*{\IEEEauthorrefmark}[1]{%
  \raisebox{0pt}[0pt][0pt]{\textsuperscript{\footnotesize #1}}%
}

\maketitle

\begin{abstract}
Threat modeling has been successfully applied to model technical threats within information systems. However, a lack of methods focusing on non-technical assets and their representation can be observed in theory and practice. Following the voices of industry practitioners, this paper explored how to model insider threats based on business process models. Hence, this study developed a novel insider threat knowledge base and a threat modeling application that leverages Business Process Modeling and Notation (BPMN). Finally, to understand how well the theoretic knowledge and its prototype translate into practice, the study conducted a real-world case study of an IT provider's business process and an experimental deployment for a real voting process. The results indicate that even without annotation, BPMN diagrams can be leveraged to automatically identify insider threats in an organization.
\end{abstract}

\begin{IEEEkeywords}
Threat Modeling, Insider Threats, Risk Management, Business Process Modeling, BPMN
\end{IEEEkeywords}

\IEEEpeerreviewmaketitle

\section{Introduction}

With the technological shift, new cybersecurity threats arise, reaffirming that enterprises face a highly active threat landscape. 
For example, a recent attack on an AI-based facial recognition system caused losses of over 77 million USD~\cite{mitre}. While empirical evidence justifies the attention to such technology and security trends,  the relevance of less technically materialized threats should not be underestimated. To share one example of exploiting processes rather than systems, the frequency of SIM swapping attacks has increased by 400\%~\cite{simswap}. 

Thus, it is critical to consider information systems wholistically, as emphasized by the National Institute of Standards and Technology (NIST), which distinguishes cybersecurity efforts into system-level, process-level, and organization-level~\cite{nistRMF}. One critical set of threats can be summarized under the term ``insider threats''. Here, a trusted person or organization intentionally or unintentionally acts as a threat actor. Such attacks are especially difficult to mitigate since technical measures only provide partial defense, and the attackers hold elevated privileges~\cite{introductionSpecialIssue}. 
In one motivating example, the Swiss government gradually rolled out a public COVID-19 certification system. 
In a wide-scale public security test, over 100 vulnerabilities were discovered and mitigated~\cite{covidTestReport}. Nevertheless, thousands of forged certificates were discovered later. The attackers did not exploit a technical flaw in the system. However, they used authorized personnel from private testing centers to issue forged certificates. In that sense, a weakness in the business process (\eg absence of a two-person rule, lack of background checks) rather than a technical vulnerability (\eg breach of authentication or authorization) was exploited~\cite{covidArticle}. 

One approach to identify threats and mitigate them at design time 
is threat modeling. In threat modeling, an abstraction of the target asset is created, enabling experts to reason about the relevant threat events. 
Threat modeling has emerged as a useful risk-based process due to its ability to work with various abstraction levels (\eg an architectural diagram or a piece of software) and different representation methods (\eg data-flow or use case diagrams)~\cite{coretm}. However, 
the application of threat modeling is often focused on technical aspects of the system. In the literature, there is a lack of methods and studies that apply threat modeling for insider threats. Furthermore, the view on the business processes encompassing different systems is often not considered, opposing the view of industry professionals who argue for capturing diverse viewpoints~\cite{manifesto}.

Due to these limitations, this paper explores the practicality of leveraging process models 
for threat modeling and therefore hypothesizes whether existing process models can serve as an input to identify relevant insider threats. 
More specifically, the hypothesis considers the usefulness of BPMN models that are processed by a security expert but not enriched by additional domain-specific annotations. Based on this automation, insider threat modeling could be included as a lightweight activity identifying specific procedural vulnerabilities around a technical information system. 
Thus, the work at hand follows the exploration of this question while presenting the following contributions. At the core, several studies describing insider threats are analyzed and ontologically combined into \1 a knowledge base. In addition to the threat events, a mapping to the BPMN elements is proposed. This knowledge base enables the development of \2 a prototypical implementation that extracts relevant insider threats from BPMN diagrams. Two experiments are conducted to assess the usefulness of this approach. First, a \3 case study, where the approach is deployed against a real-world business process of an IT service provider. In the second evaluation, \4 the approach is applied to a non-commercial setting by experimentally testing the approach's viability to provide an expert-based analysis of the Swiss voting process.

This paper is structured as follows: after surveying the state-of-the-art in Section~\ref{sec:rw}, the architecture and implementation of the approach are described in Section~\ref{sec:sol}. The evaluation of the approach is provided in Section~\ref{sec:eval}. Finally, Section~\ref{sec:conc} highlights concluding remarks and outlines future work.

\section{Related Work and Problem Statement}
\label{sec:rw}
A semi-systematic literature review targeted Google Scholar, Swisscovery, and IEEE Xplore using variations of the keyword string "insider threat modeling." After analyzing the references of these results (\ie performing snowballing), the overview, shown in \tablename{}~\ref{tab:rw} was obtained, which excluded papers not relating to security. In the table, these studies are summarized based on publication year, objective (\ie view), artifactual aspects, and methodology. Furthermore, whether a prototype is provided and insider attacks play a key role was analyzed.
Based on 14 papers, two questions were investigated. \1 How can process models, especially BPMN, be leveraged in security management, especially within the threat modeling process? \2 How does research on threat modeling consider insider threats?

Related literature on insider threat modeling can be grouped into three bags. First, numerous papers investigate the dynamics of insider threats from a psychological perspective. The focus lies on insider attacks' organizational factors and impacts. For example, \cite{introductionSpecialIssue} leverage game theory as a guiding theoretic model, yielding a simulation approach to explain the organizational factors between attacker and defender. While these studies are relevant for formulating knowledge bases, they do not directly contribute to the research questions.

The second bag of studies presents insights into threat modeling or other risk-based methods for insider threats. For example, \cite{riskAssessmentInsiderThreats, modelBasedMethodology, insiderComputerFraud}  use different methods, such as attack graphs or strategic planning methods, to demonstrate how insider threats can be integrated into the risk management of an information system. None of these solutions provide an automated threat modeling approach or focus on business process models as an existing asset to exploit.

\setlength{\tabcolsep}{3pt}
\begin{table}[b]
\centering
    \caption{Overview of the Related Work}
    \label{tab:rw}
    \begin{tabular}{@{} llllll @{}} 

    \toprule
    \textbf{\textit{Source}} & \textbf{\textit{View}} & \textbf{\textit{Aspect}} &\textbf{\textit{Methodology}} &\textbf{\textit{Prototype}} &\textbf{\textit{Insider}} \\\midrule 
    \cite{riskAssessmentInsiderThreats} 2018 & RA & IS & Several & - & yes\\
    \cite{modelBasedMethodology} 2014 & RA & IS & NIST & ADVISE & yes\\
    \cite{insiderComputerFraud} 2008 & RA & Process & TRIP & - & yes\\
    \cite{theoryInsiderthreatAssessment} 2005 & TM  & IS & Challenge Graph & - & yes\\
    \cite{processAnalysisInsiderThreats} 2014 & TM & Process & Fault Tree Analysis & Automated FTA & yes \\
    \cite{bpmnFramework} 2020 & TM & Process & SQUARE & - & no\\
    \cite{secBPMN} 2017 & TM & Process & SecBPMN & Query Engine & no \\
    \cite{autoBPMNThreatModeling} 2023 & TM & Process & ENISA, OWASP & BPMN annotator & no\\
    \cite{bpmnAttackSimulations} 2021 & TM & Process & NIST-based & coreLang & no\\
    \textit{This}  2024 & TM & Process & Knowledge Map & BPMN modeler & yes\\
    \bottomrule
\end{tabular}
RA=Risk Assessment, TM=Threat Modeling, IS=Information System
\end{table}

Finally, several solutions either cover threat modeling of insider threats or leverage BPMN as a specific abstraction of business processes to manage cyber attacks. For example, \cite{bpmnFramework} presented a methodology and prototype to check the compliance of a process through a BPMN model with previously defined security requirements. 
In addition, they provided evidence of BPMN as a useful abstraction for security modeling. However, the scope of the work did not focus on insider threats and required a previous requirements specification step. Similarly, \cite{autoBPMNThreatModeling} have relied on additional annotations to model threats based on existing frameworks provided by ENISA and OWASP. In the conducted case study, they present additional evidence on the usefulness of BPMN. However, their approach requires additional annotations and has not focused on insider threats. Regarding approaches focusing on insider attacks, \cite{theoryInsiderthreatAssessment} models insider attacks using graphs, thus not considering the process abstraction. \cite{processAnalysisInsiderThreats} is likely the most closely related study since it analyzes processes to find hazardous vulnerabilities. However, only two threats (\eg data exfiltration and sabotage) are included and for each of them a different method for the analysis is proposed. This complicates it and lacks a holistic view of all possible insider attacks. 

Based on these findings, the questions asked in the literature review can be answered. None of the approaches analyze business process artifacts to perform insider threat modeling effectively. Consequently, the research demonstrates that BPMN can be a valuable resource for security discussions. Especially \cite{autoBPMNThreatModeling} demonstrates that by using annotations, BPMN can be used for threat analysis. Since this approach was deployed in a non-commercial setting and relied on annotations, multiple avenues of research are available. In summary, the following limitations and resulting research questions drive this study.
\begin{enumerate}
    \item[\textbf{\textit{Q1}}] Is it practical to leverage existing BPMN models as an input for automated threat modeling of insider threats?
    \item[\textbf{\textit{Q2}}]Can such an approach leverage real-world BPMN models without the introduction of security-specific extensions of the visual language?
    \item[\textbf{\textit{Q3}}] How well does such an automated approach perform in commercial and non-commercial settings?
\end{enumerate}

\section{Architecture and Prototype Implementation}
\label{sec:sol}
\begin{figure*}[t]
    \centering
    \includegraphics[trim={0 .52cm 0 .39cm},clip,width=\linewidth]{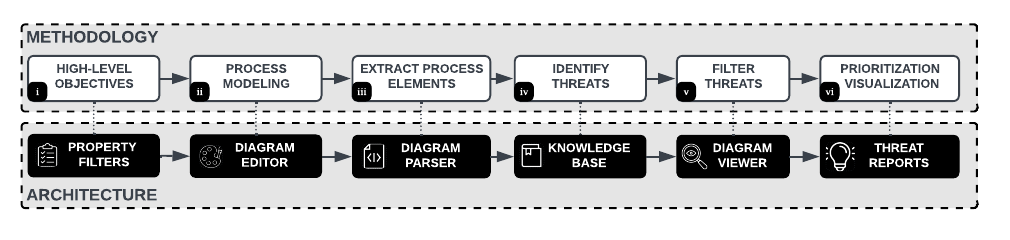}
    \caption{Methodology (top) and Architecture (bottom) of the Proposed Approach}
    \label{fig:arch}
\end{figure*}

To develop a solution that provides the means to conduct practical experiments on BPMN models for automated insider threat modeling, the architecture shown in \figurename~\ref{fig:arch} is proposed. 
A five-step process consisting of Objective Identification, Assessment, Decomposition, Threat, and Vulnerability Identification is followed as a guiding meta-model.
In the following paragraphs, each step in the meta-model is introduced to outline how the step in the proposed approach fulfills it.

%
%
%
%
\subsection{Overview}

To perform the \textit{objective identification} step, the architecture relies on the formulation of key security objectives. This involves \1 the identification of a critical process and formulating security properties through \textit{property filters} (\eg Confidentiality, Integrity). Identifying a relevant process can be informed by an implicit goal. For example, a threat modeling workshop may be tasked to model a specific system, yielding the surrounding business process. Alternatively, business metrics used in risk management such as revenue or earnings before tax and interest (EBIT)~\cite{hunziker} can be used.

To \textit{assess} and \textit{decompose} the asset and its interactions, the proposed methodology \2 involves modeling the process using BPMN. Ideally, BPMN models are already available and can be used for this security analysis. Thus, no additional annotations are mandated besides those defined in the BPMN standard, which can be modeled using a \textit{diagram editor}. This enables \3 the various interactions performed in the business process to be automatically extracted by a \textit{diagram parser}. 

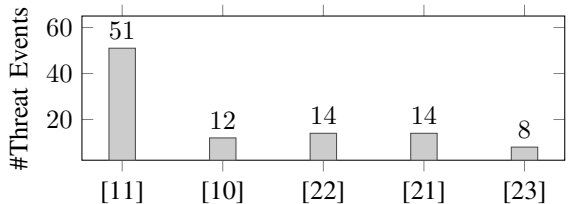
\begin{figure}[H]
    \centering
\begin{tikzpicture}  
\begin{axis}  
[  
    ybar,  
    height=3.5cm,
    width=8cm,
  cycle list={
    {fill=gray!40,draw=black!70},
    },
    ylabel={\#Threat Events}, 
    ymax=65,
    symbolic x coords={[11], [10], [22],  [21], [23]}, 
    xtick=data,  
    nodes near coords,
    nodes near coords align={vertical},  
    ]  
\addplot coordinates {([11], 51) ([10], 12) ([22], 14) ([21], 14)([23], 8) 
};
  
\end{axis}  
\end{tikzpicture}  
    \caption{Knowledge Base Sources: Insider Threats Elicited per Source}
    \label{fig:kb}
\end{figure}

\textit{Threat Identification} is the most novel aspect of the solution. To identify insider threats based on previously extracted process elements, a \textit{knowledge base} was built to synthesize the threats described in the literature. First of all, multiple sources were studied to elicit potential insider threats. After surveying several studies, 99 insider threats were elicited (see \figurename~\ref{fig:kb}) from five sources~\cite{insiderThreatSpecificationMitigation, insiderComputerFraud, combattingInsiderThreats, modelBasedMethodology, insiderThreatsInOrganizations}.  The extracted knowledge was then manually structured using a set of security principles. The commonly used principles such as \textit{Confidentiality}, \textit{Integrity}, \textit{Availability} were adopted. Furthermore, \textit{Authenticity} was included based on the reasoning in~\cite{insiderComputerFraud, securityRiskManagement}. Finally, \textit{Authenticity} was included due to its mention in \cite{networkSecurityEssentials, securityRiskManagement}.

After analyzing each threat, removing duplicates, and clustering them according to the principles, it was analyzed whether each threat could be meaningfully mapped to a BPMN element. For example, the \textit{tampering of HTTP-cookies} could be a relevant attack of an (inside) threat actor. However, comparing it against the elements in the BPMN standard, none realistically relate to it. Thus, such threat events were aggregated under the more abstract \textit{data corruption} threat.

Then, to provide a mapping to the BPMN standard, each threat event was mapped to a set containing one or more BPMN elements representing an attack entry point. For example, the threat of maliciously \textit{viewing confidential data} by an insider is relevant at different points. While it could be done at rest (\eg when an insider can access a \textit{data store}) it can be done before ingestion (\ie during a \textit{receive task}). Thus, each threat was reviewed against the BPMN elements while focusing on areas that involve human interaction (\eg activities, data objects, and message flows).

During the actual usage of the knowledge base for \4 threat identification, the knowledge base can be queried against the previously extracted elements (see \3). The resulting list of threats must then be reviewed and \5 filtered with domain knowledge. Again, experts need to consider whether a threat actually applies since BPMN models show a high degree of abstraction. For example, the knowledge base might yield a \textit{credential theft} threat. However, there may be additional technical controls in place (\eg multi-factor or biometric authentication) that are not visible from a procedural view. Thus, the resulting threat model is the subset of automatically derived threats that experts deem realistic.

The final stage of \textit{vulnerability identification} is achieved through visualization of the threat model. The goal of the threat modeling process is to identify areas of concern (\eg process steps to include additional technical or procedural controls). Since the input to the approach is already a visual representation (\ie a BPMN diagram), these areas can be highlighted. Therefore, the procedural steps in the threat model can be extracted. \cite{hunziker} argues to quantify risks -- thus, one could leverage the enterprise's data to assess the impacts of the data. Alternatively, the process steps could be color-coded to highlight the number of threat events per element within the \textit{diagram viewer}. However, this requires user awareness since a higher number of threats may not necessarily represent a higher risk.

\subsection{Web-based Implementation}
The prototype implementing the previously described methodology was developed as a web-based toolkit. Two reasons supported this design decision: first, the availability of mature BPMN diagram editors, and second, the ability to implement a completely offline data storage component. The tool follows the methodology from \figurename{}~\ref{fig:arch}. Thus, the main design decisions are elaborated per step.  

On the first page of the application, the user defines one or more security principles that are of concern. Although, in reality, every principle is likely relevant, a hint suggests starting with the most impactful one for the business process. Then, a diagram editor provided by the \textit{bpmn.io} JavaScript library~\cite{bpmnIo} is rendered, allowing the user to upload a BPMN diagram. Using the library, all the process elements are extracted. Then, a JavaScript implementation of the previously compiled knowledge base is queried directly in the browser. This set of identified threats is then visualized, grouping them per process element that represents their attack entry. 
The user then iterates through them by opening each threat in the main window.
There, each threat event is described in detail. The user can then add threats that appear relevant to the threat model, which is stored in the browser's memory. After this filtering step, the report page visualizes the selected threats. The diagram is rendered again, indicating in colors how many threats were selected per element and adding numbering to simplify the identification of each element. Furthermore, the sidebar summarizes the threats per element. Finally, since threat modeling is an iterative and collaborative endeavour~\cite{coretm}, the \textit{threat report} can be exported as a PDF, SVG, or XML file that can be loaded into the editor to repeat the process.

Due to space constraints, the solution's user interface is only illustrated in Section~\ref{sec:eval}. The solution's source code and a publicly accessible instance of the running prototype are available through~\cite{code}.

\section{Evaluations}
\label{sec:eval}
Evaluating threat modeling approaches is inherently complex. In~\cite{hunziker}, it is argued that cybersecurity (and broader information security) risks are emerging risks. For such risks, there may not be ground truth information (in this context: an exhaustive list of threats with complete information on attack probability or impact)~\cite{groundtruth}. Thus, aligned with the initial research questions, the evaluations aim to answer whether the approach can be practically applied to real-world processes. Aside from testing the input (existing BPMN models) and context (processes used by real companies), the output is assessed. Here, the focus is on the relevance of the output, \ie whether the solution can provide relevant suggestions. It must be acknowledged that the evaluations cannot demonstrate that all threats are identified since some may not even be known. Thus, a case study in a commercial setting and an expert-based field experiment within a governmental and non-commercial context was conducted.

\subsection{Case Study: IT Service Provider}
The first evaluation comprises a participatory case study described in~\cite{participatoryCaseStudy}. Following this methodology, academic and non-academic stakeholders participate in the study, representing the subject and domain experts. Importantly, the complexities and context of a real-world setting must be preserved. This is important for providing an unbiased view of the usefulness of the BPMN models as an input (\ie \textbf{\textit{RQ1}}).

Two requirements were defined for potential enterprises: the existence of formalized processes through BPMN and commercial operation. Several companies did not fit the criteria. For example, a software engineering startup did not hold formally defined processes. Four were sent a proposal to participate after surveying companies that fit the requirements. The company that enabled the case study provides IT solutions for social insurance organizations, helping them digitize their business. The business process provided by said enterprise defines the tasks a clerk in an insurance organization needs to accomplish when an insuree requests an insurance number and card.
The process involves ten tasks, two data stores, two artifacts, and an intermediate catch event to receive a message. It comprises three swim lanes, with the clerk being the central actor. Due to its size, the full process is available through~\cite{code}.

After receiving the process description, it was transformed into the target format (\ie  BPMN 2.0), and sensitive information was removed to preserve the confidentiality of technical details. Then, the process was loaded into the application, and each step (see Section~\ref{sec:sol}) was executed. Thus, the enterprise defined the key security requirements 
based on which the filtering was conducted, where each decision was documented for later review by the domain experts. Finally, the resulting threat model and its creation were presented to the company. 

\begin{figure*}
    \centering
    \includegraphics[width=\linewidth, trim={0 7cm 0cm .55cm},clip]{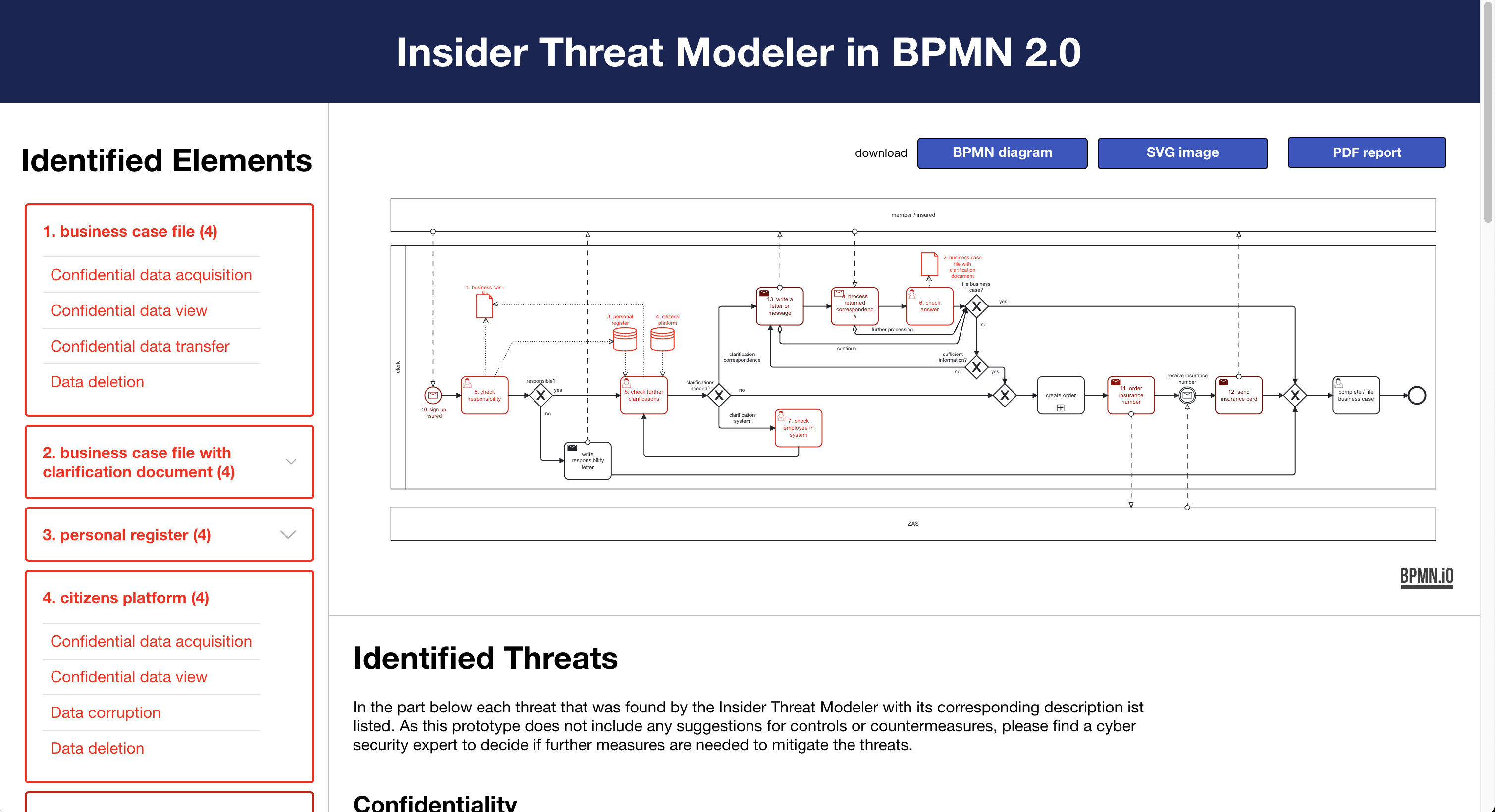}
    \caption{Report Page of the Threat Model Conducted for the IT Service Provider}
    \label{fig:casestudy-results}
\end{figure*}

\figurename{}~\ref{fig:casestudy-results} illustrates the final report page of the case study -- highlighting the threat overview, which is grouped by asset. Additional information on each threat is rendered below the process diagram. The threat model (see \tablename~\ref{tab:threatReportCaseStudy}) comprised seven key threats (\eg Data Acquisition, Data View) and thirteen assets (\ie process steps or artifacts found in the process). The most critical areas were the \textit{Case File}, the \textit{Personal Register}, and the \textit{Citizen Platform}. Thus, these key assets are highlighted, and the number of applicable threats is displayed.

\begin{table}[t]
    \caption{Threat Report of Case Study}
    \label{tab:threatReportCaseStudy}
    \begin{tabular}{@{} lll @{}}    
    \toprule
    \textbf{\textit{Asset}} & \textbf{\textit{Type}} & \textbf{\textit{Insider Threats}} \\\midrule
    Business Case File & Data Object  & DA, DV, DT, DD
    \\
    Case File, Clarification Document & Data Object & DA, DV, DT, DD \\
    Personal Register & Data Store & DA, DV, DC, DD
    \\
    Citizens Platform & Data Store & DA, DV, DC, DD 
    \\
    Check Further Clarifications & User Task & DC, SC, DD 
    \\
    Check Answer & User Task & DC, SC, DD
    \\
    Check Employee in System & User Task & DC, SC, DD\\
    Check Responsibility & User Task & DC, SC, DD
    \\
    Process Return Correspondence & Message Receive & DV, MI
    \\
    Sign Up Insuree & Message Receive & DV, MI
    \\
    Order Insurance Number & Message Send & DT, DC
    \\
    Send Insurance Card & Message Send & DT
    \\
    Notification Letter & Message Send & DC \\
     \bottomrule
\end{tabular}
     DA=Data Acquisition, DV=Data View, DT=Data Transfer, DC=Data Corruption, DD=Data Deletion, SC= System Control Manipulation, MI=Malware Installation
\end{table}

After the methodology and results were presented, a feedback session was held involving the enterprise's security manager (CISO), its deputy, and the process architect. The results of the semi-structured interview are summarized subsequently.

\begin{itemize}
    \item \textbf{True positives: are the threats relevant and known?}

    All discovered threats were considered relevant by the experts. Furthermore, the experts agreed on their importance since the confidentiality of the information (digital and physical) is highly sensitive. The company already considered these threats in its threat model and implemented security controls. 
    
    \item \textbf{False negatives: are there threats that were not found?}

    The security expert indicated that injection, privilege escalation, and social engineering attacks could be relevant in addition to the elicited threats.
    
    \item \textbf{Are there countermeasures or controls already implemented where the critical elements were identified?} 

    According to the CISO, the organizational level adopted controls such as employee education, fine-grained access control, and confidentiality agreements. Furthermore, an audit trail ensures that any modifications can be traced. For critical decision-making steps, the four-eyes principle is integrated. 
    From a technical perspective, \textit{Data Deletion} is mitigated through backup and archiving procedures. \textit{Data Corruption} through logging and \textit{Malware Injection} through filtering and sanitization.

    \item \textbf{Are there assets (\ie process elements), where other threats than the ones identified could be relevant?}

    The CISO expressed that the database could be attacked through a denial-of-service attack using a complex database query.
    

\end{itemize}

Based on this interaction, it could be concluded that the approach can suggest relevant threats by relying on the BPMN model. However, the output will not exhaustively represent all feasible attacks. This is important in relation to several biases commonly found in risk management~\cite{hunziker}. Thus, communicating the results requires careful consideration. 

The company has already considered all the threats through its security program. While one could conclude that the solution could not stipulate "novel threats," another interpretation is that the enterprise's program is already quite mature. For example, procedural controls such as the four-eyes principle were not found in the motivating example shared at the beginning of the paper~\cite{covidArticle}.
In that sense, the presence of control mechanisms indicates that the provided threats are relevant enough to invest in their mitigation.



\subsection{Field Experiment: Remote Voting}
In the second evaluation, a non-commercial setting was considered, which exhibits different characteristics. For example, \cite{egov}~argue that decision-making, which is needed to secure governmental information systems, is much more difficult due to the decentralized nature of authority.

This experiment considered an established but not formalized business process. More specifically, the process of remote postal voting described in the security analysis~\cite{killer2019swiss} was examined. In the paper, the wholistic voting procedure spanning several processes and actors is described. For this experiment, only the process of casting a vote through different channels, storing, and tallying the votes is used. The experiment investigates whether a security professional can leverage the application to formalize the process using BPMN and conduct a meaningful insider threat identification process. The expert in the study has several years of technical work experience in security engineering but no domain knowledge of voting systems and processes. The models created during the experiment are available in~\cite{code}.

First, the security expert was provided with the description from~\cite{killer2019swiss} and the instruction to focus on the \textit{casting}, \textit{storage}, and \textit{tallying} aspects. After opening the tool, the security provider focused on \textit{Integrity} and \textit{Confidentiality}  as key objectives. Then, the security expert formalized the process using the built-in BPMN editor. The expert modeled the different voting mechanisms through three swim lanes, representing the different actors (\ie voters, postal service, municipality). In the third stage, the expert navigated the threats identified by the platform, performing the filtering that resulted in the final threat model. Thus, seven unique threats were analyzed, some affecting several steps. After the filtering step, the threat model comprised 16 threat events affecting 10 
assets (\ie process steps). If no filtering by means of security goal definition or threat applicability analysis had been done, the result would have comprised 52 threats. The experiment lasted two hours, including familiarization with the application's usage, understanding the voting domain, and actual execution.

Once the threat model was created, the author of~\cite{killer2019swiss} was contacted to assess the resulting threat model. Since there are no ground-truth threat models, the author's expertise is used for comparison since the author is considered an expert in this field. The resulting threat report was shared as a PDF, and the process model contained the automatically generated color highlighting. After 15 minutes of studying, the following findings were obtained using a semi-formal interview:

\begin{itemize}
    \item \textbf{True positives: are the threats relevant and known?}
      According to the author, all threats identified in the report are highly relevant to the scope of the remote postal voting process. Furthermore, the threat model highlighted the highest number of threats for the areas deemed most sensitive by the author. Here, the tool explicitly highlighted two areas: \1 the temporary storage between receiving votes and starting the tallying process was highlighted, and \2 the store-and-forward processing in the postal service. A malicious actor could view, corrupt, or destroy the material. Interestingly and unbeknownst to the security expert, this threat has previously occurred. In 2020~\cite{fallTG} a local election worker (\ie inside actor) destroyed voting material, causing a change in the results.

    \item \textbf{False negatives: are there threats that were not found?}
      The author highlighted multiple false negatives. Most importantly, artefact corruption (\ie modifying voting material) would be relevant at almost all steps in the process. However, the threat model indicated it in eight out of ten steps. Furthermore, deletion attacks were missing (\ie destroying or hiding voting material). This is attributed to the threat's association with \textit{Availability} in the knowledge base. Hence, it was falsely excluded. Finally, the threat of introducing malicious code could be more prevalent, as modeled by the security expert.

    \item \textbf{Are there countermeasures or controls already implemented where the critical elements were identified?} 
      The critical threats modeled by the security expert were mostly focused on the postal service and the municipalities' storage facilities. According to the author's research, the postal service already includes countermeasures and has been audited against several related standards (\eg ISO 27001, ISO 22301). However, the author stated that the key area indicated by the threat model (\ie municipal vote storage) is not necessarily covered by physical security controls. This is magnified by the fragmented view since every municipality operates differently and since no central coordinator is overviewing the process. 

    \item \textbf{Are there assets (\textit{i.e.} process elements), where other threats than the ones identified could be relevant?}
According to the author, the BPMN representation includes all relevant procedural assets. Importantly, this does not prove that it actually reflects the real-world process since the input process diagram is already an abstraction.
\end{itemize}

Based on these questions and the additional views expressed by the author, several conclusions can be drawn from the experiment. First, the prototype was able to identify relevant threats. In a different setting, for example, if the prototype was applied on the municipal level, discussions on the security of the process could be stipulated. The sensitive area of vote storage could lead to a discussion on additional physical and digital control mechanisms (\eg four-eyes principle, two-key lockbox, statistical plausibility tests). 

However, the application of the prototype also led to the discovery of multiple limitations. \1 The time required to adopt the tool and formalize undocumented processes may be too costly and only efficient for critical procedures.  
\2 Understanding the threat reports requires some interpretation since the descriptions are generic and not domain-specific. For example, in the threat model, a key threat is that the \textit{voting proof} could be stolen from the envelopes. In the PDF, this proof is simply referred to as \textit{credential} and thus requires some time for interpretation. The key limitation expressed by the author is \3 that in a non-commercial setting, the initial prioritization is not as sensible. 
For example, one could argue that integrity (\eg forging votes) is more impactful than confidentiality (\eg losing ballot privacy). However, in practice, different stakeholders may hold different views. Thus, the author argued that in such a setting, everything is relevant.

In addition to these findings related to the overall usage of the tool and its approach, the author suggested usability improvements (\eg numbering the threats, visualizing them within the diagram).


\subsection{Synthesis and Comparison with Related Work}
At this point, the study's initial research questions are reviewed, and findings are highlighted and compared to related work. Regarding \textit{\textbf{Q1}}, the real-world deployments 
demonstrated that the underlying approach could successfully identify relevant threats. Furthermore, the experiments indicated that the output can foster actionable discussions. In comparison, \cite{processAnalysisInsiderThreats} argued that process abstraction is suitable to model insider threats. However, this approach was only illustrated and not deployed for a real-world scenario. Similarly, \cite{bpmnAttackSimulations} performed process-based insider threat analysis; however, 
the resulting model was not reviewed 
by a user. Thus, although one must be cautious about generalizing based on case studies, evaluating this work provides insights into the approach's real-world practicality.
Overall, the key limitation of this aspect is the proper communication in the domain's context. 

With respect to \textit{\textbf{Q2}}, the prototype worked well with existing process descriptions and when onboarding a security expert to create a new one. \cite{autoBPMNThreatModeling} demonstrated that through annotations introducing technical system details (\eg log level, message size), it is possible to identify threats automatically. In this work, no modifications from the BPMN standard were required.  Nevertheless, an overarching concern discovered in such process-based techniques lies in the availability of such data -- further research is needed to answer whether it is cost-effective to create new models solely for this purpose.

Overall, the approach appears effective in identifying \textit{relevant} threats (\textit{\textbf{Q3}}). Nevertheless, filtering threats may be less practical in non-commercial settings. In that sense, it questions related studies that have focused only on governmental domains~\cite{autoBPMNThreatModeling}. A key limitation of the work at hand is the potential presence of false negatives since the filtering may exclude certain threats. 
From a methodological perspective, experiments conducted in this work carry a participatory character and are thus subject to the limitations described by~\cite{participatoryCaseStudy}. In addition, while case study research enables collecting rich experience on the methodology, generalization is an issue due to the low number of samples.

\section{Summary and Future Work}
\label{sec:conc}
This paper proposed a threat modeling approach to identify insider threats using BPMN process models. The approach and its implementation cover eliciting security requirements, modeling, and extracting information from existing processes. Then, the threat identification phase leverages a bespoke knowledge base that synthesizes findings from several studies and maps them to BPMN elements. After filtering threats, a simple prioritization is suggested, and threats are visualized in the procedural context. The experiments (\ie a case study and field experiment) conducted using the platform suggest that existing BPMN models can serve to identify relevant threats using the approach. This includes real-world settings -- even when considering commercial and non-commercial settings. Thus, it can be concluded that BPMN is an effective tool for understanding the procedural context surrounding information systems and how to secure it using threat modeling. Based on the deployment against real-world scenarios, it is suggested that procedural threat modeling can serve as a critical abstraction to model the boundaries between systems and operators.

The key limitations include the effort required by analysts, the presence of false negatives, and the prioritizing of threats based on security requirements in non-commercial settings. Thus, future work will include diversifying the experiments and applying other risk-based prioritization methods (\eg quantitative simulations, and business impact analysis).

\section*{Acknowledgment}

This work was supported by the University of Zürich UZH. Furthermore, the authors thank Christian Killer for his time.

\bibliographystyle{IEEEtran}  
\balance
\bibliography{references}
\noindent \small{\\All links above were last accessed on \today{}.}

\end{document}